\documentclass{aa}
\usepackage{caption}
\usepackage{graphicx}
\usepackage{txfonts}
\usepackage{lipsum}
\usepackage{subcaption}         
\usepackage{lscape}             
\usepackage{placeins}           
\usepackage{epstopdf}           
\usepackage{natbib}
\usepackage{longtable}
\usepackage{booktabs}
\bibpunct{(}{)}{;}{a}{}{,} 

\usepackage{caption}
\usepackage{color}
\usepackage{graphicx}

\definecolor{ora}{rgb}{1,0.5,0} 

\begin{document}

   \title{The invisible threat}

   \subtitle{Assessing the collisional hazard posed by the undiscovered Venus co-orbital \\ asteroids}

   \author{V. Carruba \inst{1,2}, R. Sfair\inst{3,1}, R. A. N.  Araujo\inst{1}, O. C. Winter\inst{1}, D. C. Mour\~{a}o\inst{1}, S. Di Ruzza\inst{4}, S. Aljbaae\inst{5}, G. Carit\'{a}\inst{5}, \\ R. C. Domingos\inst{6}, A. A. Alves\inst{1}.}
   \authorrunning{Carruba et al.} 
   \institute{S\~{a}o Paulo State University (UNESP), School of Engineering and Sciences, Guaratinguet\'{a}, SP, 12516-410, Brazil.
             \email{valerio.carruba@unesp.br}
             \and Laborat\'orio Interinstitucional de e-Astronomia, RJ 20765-000, Brazil.
              \and LIRA, Observatoire de Paris, Université PSL, Sorbonne Université, Université Paris Cité, CY Cergy Paris Université, CNRS,  92190 Meudon, France.
              \and Dept. of Mathematics and Informatics, Univ. of Palermo, Palermo, 90123, Italy.
             \and National Space Research Institute (INPE), Postgraduate Division, S\~{a}o Jos\'e dos Campos, SP 12227-310, Brazil.
             \and S\~{a}o Paulo State University (UNESP), School of Engineering, S\~{a}o Jo\~{a}o da Boa Vista, SP, 13876-750, Brazil.
            }

   \date{Received February 28, 2025}

 
  \abstract
   {Currently, 20 co-orbital asteroids of Venus are known, with only one 
with an eccentricity below 0.38.  This is most likely caused by observational
biases since asteroids with larger eccentricities may approach the Earth and 
are easier to detect. }
   {We aim to assess the possible threat that the
yet undetected population of Venus co-orbitals may pose to Earth, and investigate
   their detectability from Earth and space observatories.} 
   {We used semi-analytical models of the 1:1 mean-motion resonance with Venus and numerical simulations to monitor close encounters with Earth on several co-orbital cycles.  
   {We analyzed observability windows and brightness variations for potential Venus 
   co-orbitals as viewed from ground-based telescopes to assess their future 
   detection feasibility with next-generation survey capabilities.}}
   {There is a range of orbits with $e< 0.38$, larger at lower inclinations, for which Venus'co-orbitals can pose a collisional hazard to Earth.}
   {{Current ground-based observations are constrained by periodic observing windows and solar elongation limitations, though the Rubin Observatory may detect some of these objects during favorable configurations. Space missions based on Venus'orbits may be instrumental in detecting Venus'co-orbitals at low eccentricities.}}


   \maketitle

\section{Introduction}
\label{sec: intro}
\nolinenumbers
Venus co-orbital asteroids are objects that are in a 1:1 mean-motion resonance with
Venus.  At low eccentricities, co-orbital configurations include tadpole orbits (T) around the triangular $L_4$ and $L_5$ Lagrangian equilibrium points (TL4 and TL5), and horseshoe orbits (H) \citep{1999ssd..book.....M}.  At higher eccentricities other types of orbits, like Retrograde Satellites (RS), H-RS compound orbits, which are mergers of RS and H, or H and RS, and T-RS compound orbits, resulting from a combination between Tadpoles and RS, as well as transitions between any of these configurations, are also possible (\citep{Namouni1999a, 2000Icar..144....1C, Morais2006}, see also Section~(\ref{sec: ham_11}) for a discussion of these concepts).  The known Venus co-orbital asteroids typically spend $12000\pm6000$ years alternating between different co-orbital configurations, until transitioning to circulating or passing orbits in what \citet{Carruba2025} defined as a co-orbital cycle. Some asteroids may pass through several co-orbital cycles, and remain in circulating orbits in-between these cycles.  These types of orbits are defined as ``dormant'' in \citet{Carruba2025}.

Twenty co-orbital asteroids of Venus are currently known (see \citep{2024RNAAS...8..213C, pan2024attemptbuilddynamicalcatalog} and references therein).  Co-orbital status protects these asteroids from close approaches to Venus, but it does not protect them from encountering Earth \citep{2000Icar..144....1C}. A Potentially Hazardous Asteroid ($PHA$) must have a minimum diameter of about 140 meters and come within 0.05 astronomical units (au) of Earth's orbit. The maximum absolute magnitude ($H$) for an asteroid to be classified as a PHA is typically around 22 \citep{PHA_task_force_2000}. \citet{Carruba2025} identified six co-orbital asteroids that can become $PHA$s in the next $\simeq 12000$ years, with three asteroids, 2020 SB, 524522, and 2020 CL1, having a Minimum Orbital Intersection Distance (MOID), which is the distance between the closest points of two objects' orbits, with Earth of less than 0.0005 au. I.  Assuming a geometric albedo $p_V$ of 0.12 and using the absolute magnitudes of these objects and standard methods to estimate asteroid sizes \citep{bowell1989application}, their diameters would be 330, 300, and 390 meters, respectively.  Based on results from an impact simulator from \citet{Collins2005}, assuming a density of 1200 $kg/m^3$, typical of the common C-complex asteroids \citep{CARRY201298}, an impact angle of $45^{\circ}$, a velocity equal to the escape velocity from Earth of 11.2 km/s, and impacts on sedimentary rocks, these objects could form craters with diameters from 2.2 to 3.4 Km, and release energies at impact from $1.5$ to $ 4.1 \times 10^2$ Megatons TNT, which is more than enough to destroy large cities. This corresponds to a level 8 in the Torino scale \citep{binzel1998torino}, i.e., a collision capable of causing localized destruction for an impact over land or possibly a tsunami if close offshore.

Currently, only one of the known Venus'co-orbitals has a value of eccentricity ($e$) slightly lower than 0.38, which is the limit for which the apocenter of the orbit of a Venus co-orbital is equal to the Earth pericenter.  This phenomenon is not limited to just the co-orbitals of Venus, but is a characteristic of all asteroids in the region, as discussed in Section~(\ref{sec: neomod3}).  Two possible mechanisms have been proposed to explain the lack of observations of asteroids at lower eccentricities.  In the dynamical scenario, NEAs are created disproportionally at higher eccentricities, while in the observational bias model \citep{pan2024attemptbuilddynamicalcatalog, 2023Icar..39015330D}, a large population of NEAs exists near Venus at low $e$, but it is not detected because of biases in observational data, that tend to privilege the discovery of faint objects that came closer to Earth and whose orbit could be external to the Earth’s one.  Results from recent models of the production of NEAs from the main belt and other dynamical source regions \citep{2024Icar..41716110N} show that there is no observed bias for producing low-$e$ asteroids near Venus.  This means that it is very likely that a large population of Venus co-orbital asteroids at $e < 0.38$ may exist, but it is yet undetected.

In this work, we aim to assess the collisional hazard posed by undetected Venus Trojans.  For this purpose, we use the semi-analytical model by \citet{pan2024attemptbuilddynamicalcatalog} to set up initial conditions for long-term simulations of Venus'co-orbitals, in a grid of initial conditions in the $(e, inc)$ plane, with $inc$ being the inclination of the orbital plane of the orbits with respect to the orbital plane of Venus, that includes the possible undetected population of Venus co-orbital asteroids (see Sect.~(\ref{sec: ham_11})).  Close encounters with Earth are then monitored using the method described in \citet{Carruba2025}, to provide information on the orbital location of the most dangerous undetected co-orbitals (see Sect.~(\ref{sec: close_enc})). Based on this analysis, in Sect.~(\ref{sec: observ_earth}) we discuss the feasibility of observing these asteroids from Earth, with an emphasis on observations from the Vera C. Rubin Observatory, or Large Synoptic Survey Telescope (LSST), while in Sect.~(\ref{sec: observ_venus}) we review current proposals for space mission to observe asteroids from orbits near Venus.  Finally, in Sect.~(\ref{sec: concl}) we present our conclusions.

\section{The expected orbital distribution of NEAs near Venus: NEOMOD3 results}
\label{sec: neomod3}

\begin{figure*}
  \centering
  \begin{minipage}[c]{0.45\textwidth}
    \centering \includegraphics[width=3.in]{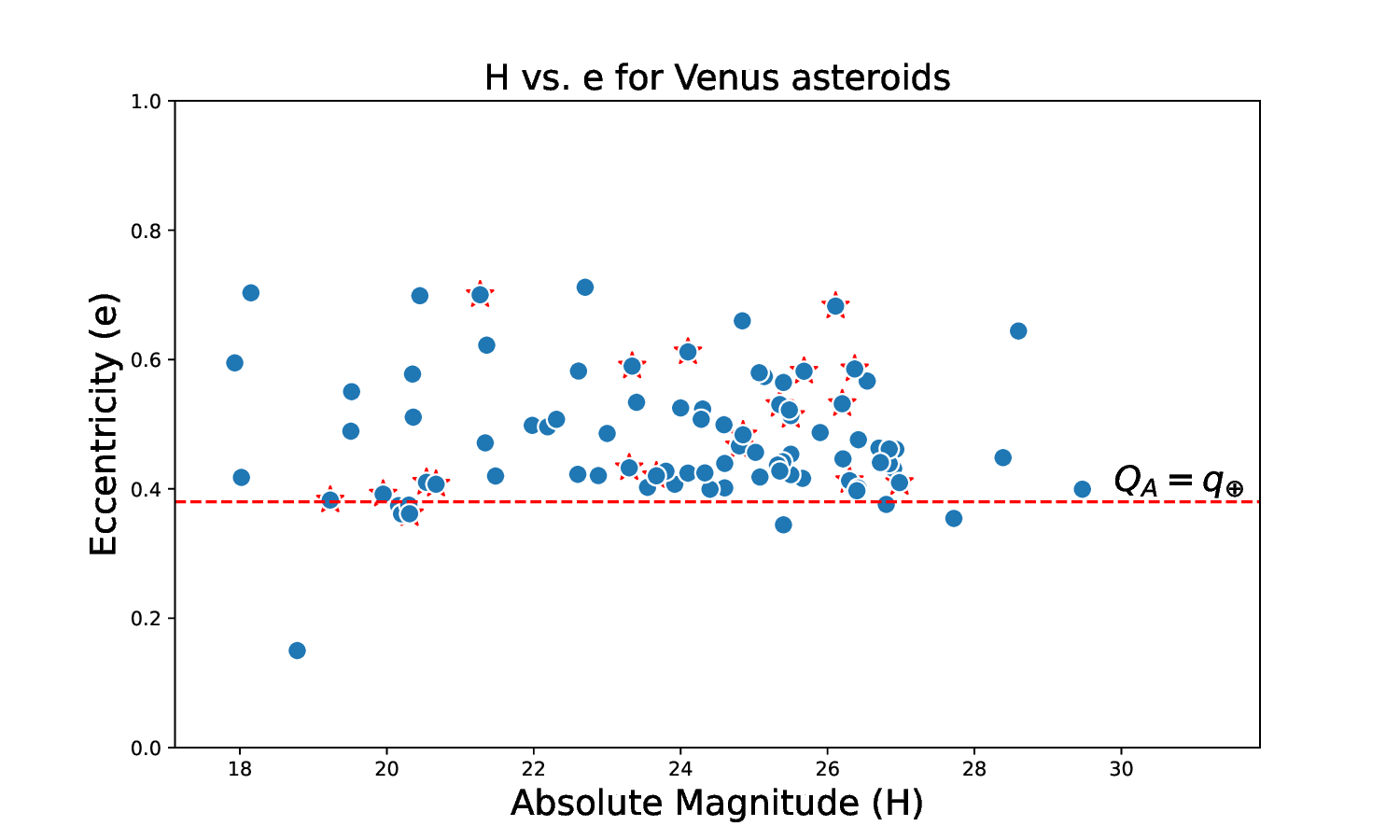}
  \end{minipage}%
  \begin{minipage}[c]{0.45\textwidth}
    \centering \includegraphics[width=3.in]{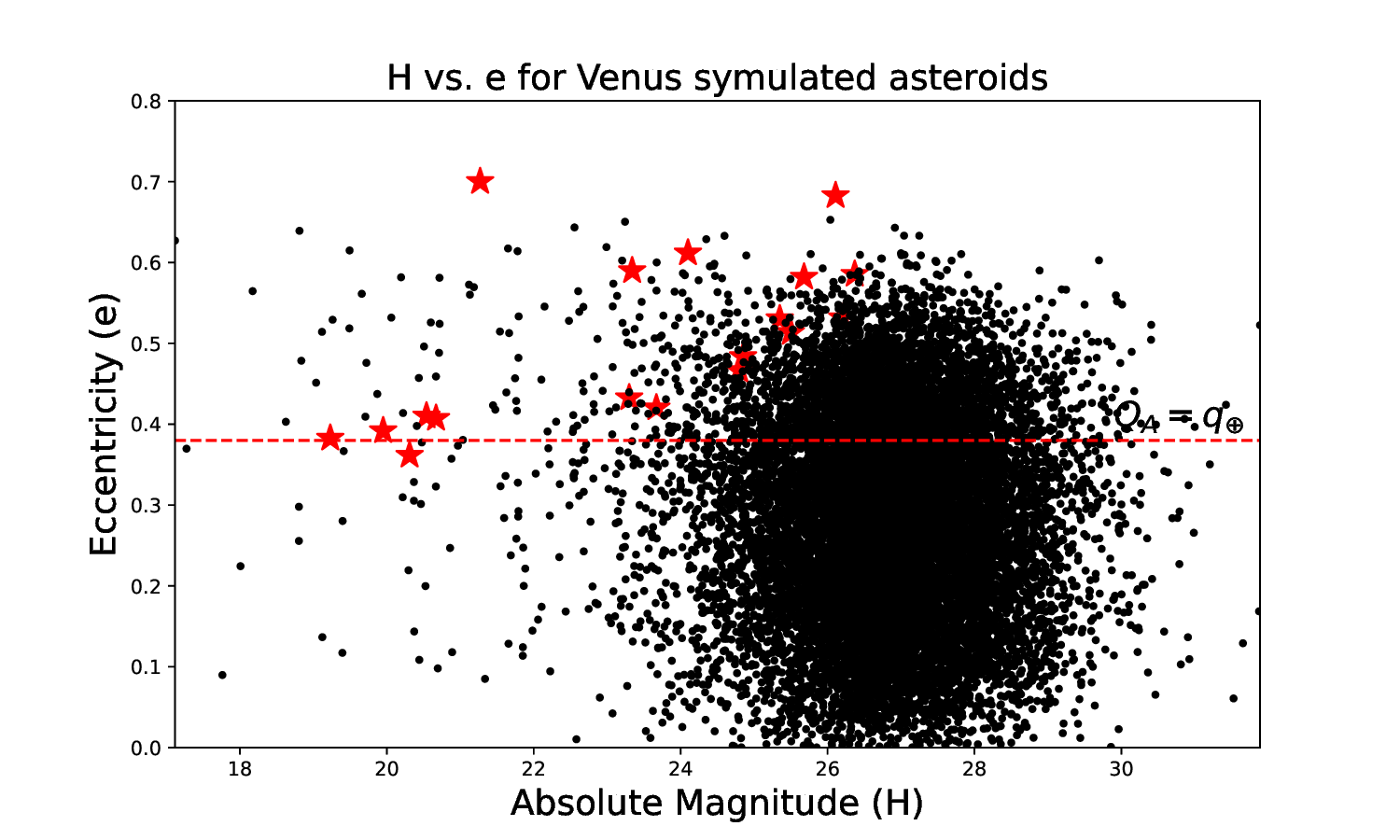}
  \end{minipage}
  \caption{Left panel: the distribution in the $(H,e)$ plane of known asteroids near Venus (blue full circles) and its co-orbital asteroids (red stars).  The horizontal dashed line shows the limit for which the apocenter of Venus' co-orbitals ($Q_A$) can be equal to the Earth's pericenter ($q_{\oplus}$).  Right panel: the distribution of 14382 simulated NEAs obtained from the NEOMOD3 model (black dots).  The other symbols have the same meaning as in the left panel.}
  \label{Fig: neomod3}
\end{figure*}

As discussed in Section~(\ref{sec: intro}), all but one of the currently known
Venus'co-orbital asteroids have eccentricities greater than 0.38.  This trend also applies to all other known asteroids near Venus. In the left panel of
Fig.~(\ref{Fig: neomod3}) we
show the distribution of Venus' co-orbitals (red stars) and all the asteroids
in the $a$ range (width of 0.032 au, between 0.707 and 0.739 au) for which \citet{pan2024attemptbuilddynamicalcatalog} assessed that co-orbital behavior is possible (blue full dots).  Only six asteroids have eccentricity values of less than 0.38 in the region. Since co-orbital asteroids of Venus are transient, and stay in co-orbital cycles of $12000\pm6000$ years \citep{Carruba2025}, one could postulate if the lack of objects with lower $e$ could be caused by dynamical mechanisms or by observational biases.

The first hypothesis can be quickly ruled out using modern models of the expected population of NEAs in the same $a$ range. Here, we use the NEOMOD3
model of \citet{2024Icar..41716110N} to simulate the absolute magnitude and orbital distribution of asteroids with diameters between 0.01 and 10 km in the area. NEOMOD is an orbital and absolute magnitude model of Near-Earth Objects (NEOs). The model was developed following the methodology of previous studies and was improved when possible. Massive numerical integrations have been performed for orbits of asteroids escaping from eleven major sources in the asteroid belt. Comets were also included as the twelfth source. Integrations were used to calculate probability density functions (PDFs) that define the orbital distribution of NEOs from each source.  In the area of interest, we find 14382 simulated asteroids, whose distribution in the $(H,e)$ plane is shown in the right panel of Fig.~(\ref{Fig: neomod3}).  There is no preference for producing NEOs with higher eccentricities, and the actual distribution in $e$ is very uniform, with a large number of objects expected to be found with $e< 0.38$.

Having ruled out dynamically based reasons, observational biases should be the most reasonable explanation for the lack of known co-orbitals at $e< 0.38$. The known co-orbital asteroids of Venus are faint objects with absolute magnitudes of 19 or more, that are only observable from Earth with large telescopes, at very high azimuth angles, and just for small periods after sunset or before sunrise.  Objects that come closer to Earth have higher apparent magnitudes and are easier to detect.  This does not mean that lower eccentricity co-orbital asteroids do not exist.  As discussed in \citet{Carruba2025}, some of Venus' co-orbital could become Potential Hazardous Asteroids (PHAs) in a few hundred years.  In this work, we will assess the possible collisional hazard that a population of yet undetected small eccentricity Venus co-orbitals can pose to Earth.

\section{Hamiltonian model of the 1:1 resonance}
\label{sec: ham_11}

\citet{pan2024attemptbuilddynamicalcatalog} recently used a semi-analytical approach based on the model developed by \citep{2020CeMDA.132....9G} to study co-orbital motion in 1:1 mean-motion resonances across the Solar System. In this section, we describe this model and apply it to the case of Venus' co-orbitals. In general, this model, compared to analytical ones, is more straightforward and only provides instantaneous features of the dynamics on the assumption that the orbital elements $(e, inc, \omega, \Omega)$ of the asteroid are fixed, being $e$ the eccentricity, $inc$ the inclination, $\omega$ the argument of the pericenter, $\Omega$ the longitude of the ascending node. Moreover, it allows us to understand the resonance structure as the orbital elements evolve secularly but we are unable to determine and predict the evolution of the dynamics on a long-term scale.  This kind of analysis will require numerical simulations, to be discussed in Sect.~(\ref{sec: close_enc}).
The Hamiltonian describing the motion of a particle in 1:1 mean-motion resonance with a planet around a star is given by

\begin{equation}
H(a,\sigma) = - \frac{\mu}{2a}-n_p \sqrt{\mu a} -R(a_0, \sigma),
\label{eq: Pan_Hamiltonian}
\end{equation}

\noindent with $\mu = GM_{\star}$, where $G$ is the gravitational constant, $M_{\star}$ is the Sun mass, $a$ is the particle's semi-major axis, $n_p$ is the planet's mean motion, and $a_0$ is the resonance's nominal value, namely, the mean semi-major axis of the planet. $\sigma = \lambda - \lambda_2$ being is resonant angle for the 1:1 mean-motion resonance with Venus. The first term of the Hamiltonian is the Keplerian term that governs the motion of the asteroid around the Sun. The second term is due to the extended phase space and the third one, given by the function $R$, represents the resonant disturbing function due to the perturbation of the planet on the asteroid motion. The function $R$ is numerically determined and computed using fixed orbital elements $(e, inc, \omega, \Omega)$, assuming that they are constant during one or a few orbital periods. 

 The model also offers several resonance characteristics, such as the width and the center of libration, and the period for small amplitude librations. Since most of the known Venus' co-orbitals are located at inclinations ($inc$) lower than $45^{\circ}$ and $e < 0.60$ \citep{2024RNAAS...8..213C}, here we apply \citep{pan2024attemptbuilddynamicalcatalog} model to obtain information about the relevant dynamics in this orbital region.

\begin{figure*}
  \centering
  \begin{minipage}[c]{0.45\textwidth}
    \centering \includegraphics[width=2.3in]{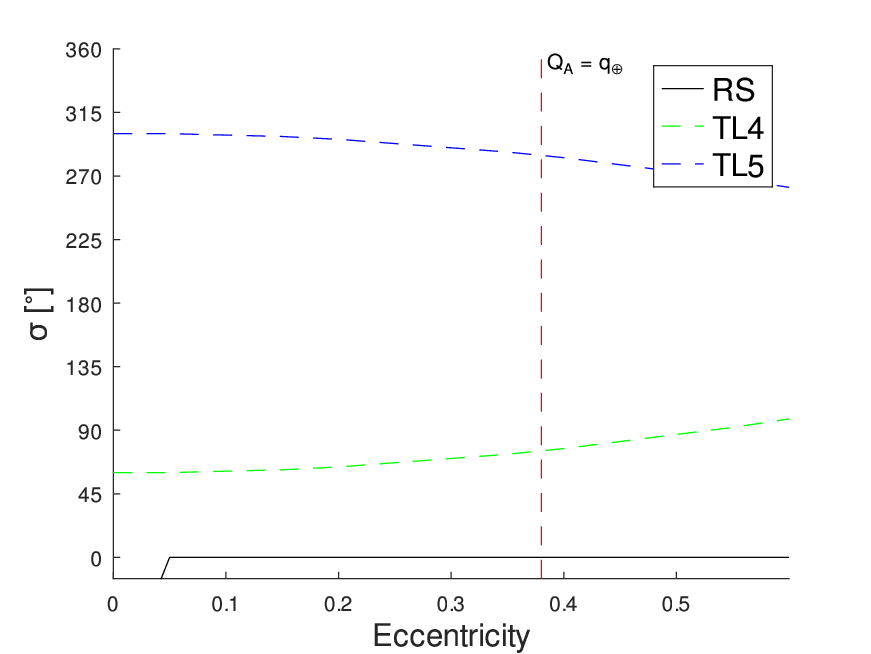}
  \end{minipage}%
  \begin{minipage}[c]{0.45\textwidth}
    \centering \includegraphics[width=2.3in]{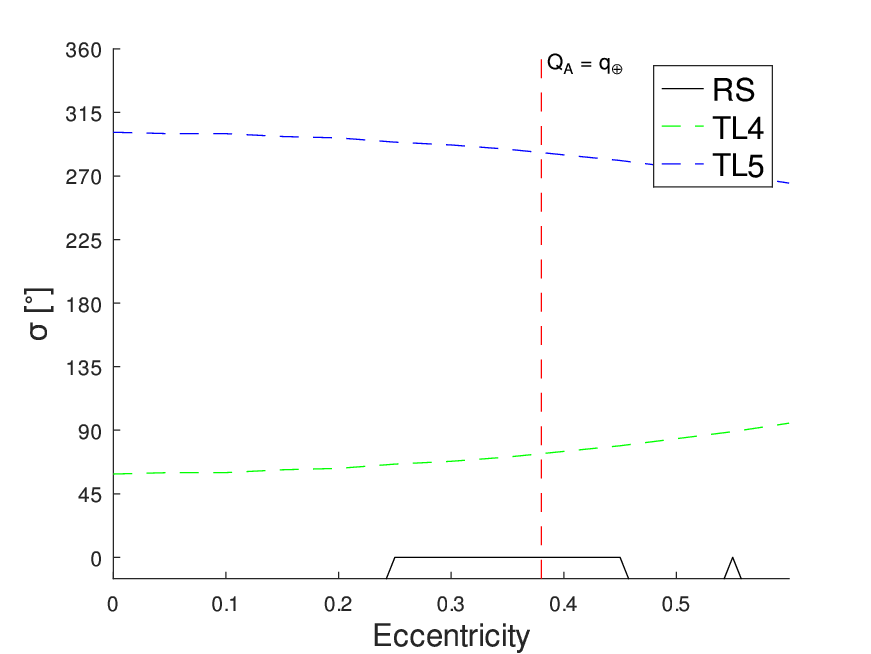}
  \end{minipage}
  \begin{minipage}[c]{0.45\textwidth}
    \centering \includegraphics[width=2.3in]{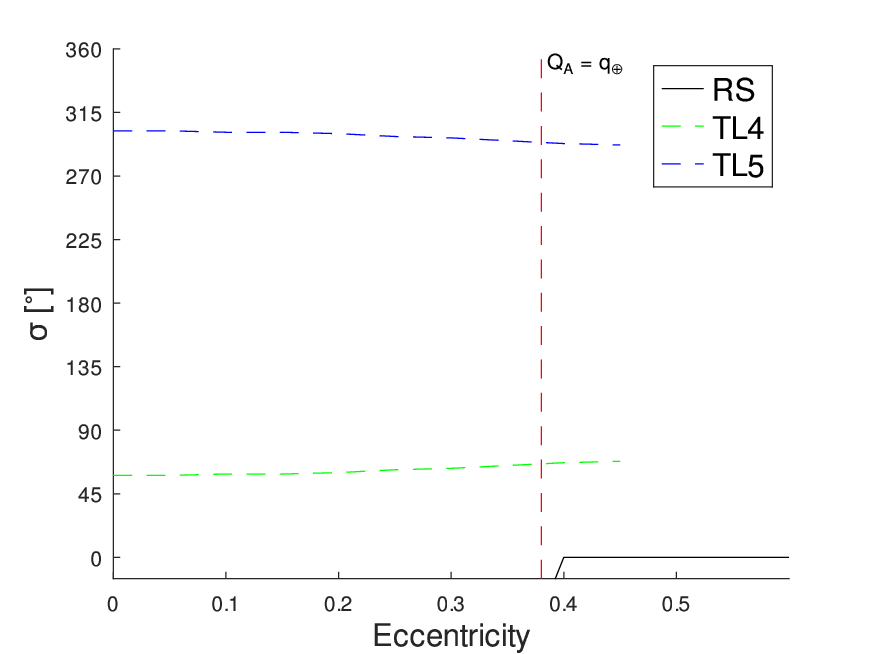}
  \end{minipage}%
  \begin{minipage}[c]{0.45\textwidth}
    \centering \includegraphics[width=2.3in]{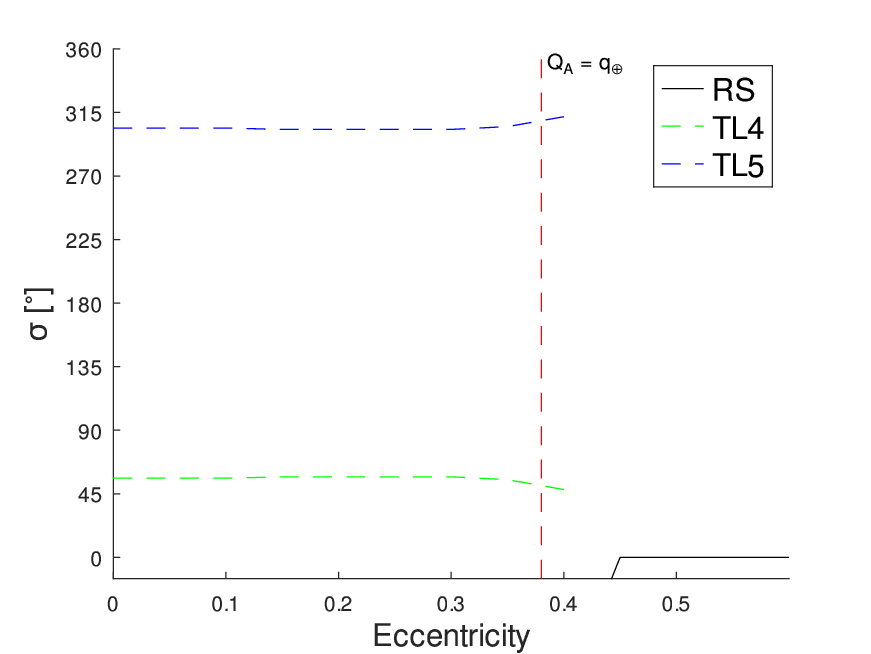}
  \end{minipage}
  \caption{Location of the equilibrium points in the $(e,\sigma)$ plane for
  $inc = 0^{\circ}$ (top left panel), $inc = 15^{\circ}$ (top right panel), $inc = 30^{\circ}$ (bottom left panel), and  $inc = 45^{\circ}$ (bottom right panel).  The vertical dashed line displays $e=0.38$, the level for which $Q_A = q_{\oplus}$.}
  \label{Fig: eq_points}
\end{figure*}

Figure~(\ref{Fig: eq_points}) displays the location of the equilibrium points, corresponding to the centers of librations, in the $(e,\sigma)$ plane for four values of $inc$, where $\lambda$ and $\lambda_2$ are the mean longitude of the asteroid and Venus (designated by the suffix 2), respectively.  At $inc = 0^{\circ}$ the two equilibrium points
at the Lagrangian points $L_4$ and $L_5$ are initially at $60^{\circ}$ and $300^{\circ}$.
As initially observed by \citep{Namouni1999a}, at higher eccentricities they
move closer together.  The equilibrium point at $0^{\circ}$ is not initially
observed at $e = 0$, but it appears for greater eccentricities.
At $inc = 15^{\circ}$, the behavior of the $L_4$ and $L_5$ Lagrangian point is similar
to the case for $inc = 0^{\circ}$, but the $0^{\circ}$ equilibrium point
is only observable for a smaller $e$ range, between 0.25 and 0.45.  A similar behavior can be perceived at $inc = 30^{\circ}$, with the $0^{\circ}$ equilibrium point now limited to a range in $e$ between 0.4 and 0.6.  Finally, for
$inc = 45^{\circ}$, the $L_4$ and $L_5$ equilibrium point disappear at $e = 0.4$,
and the $0^{\circ}$ equilibrium point only appears for $e > 0.45$.

\begin{figure*}
  \centering
  \begin{minipage}[c]{0.45\textwidth}
    \centering \includegraphics[width=3in]{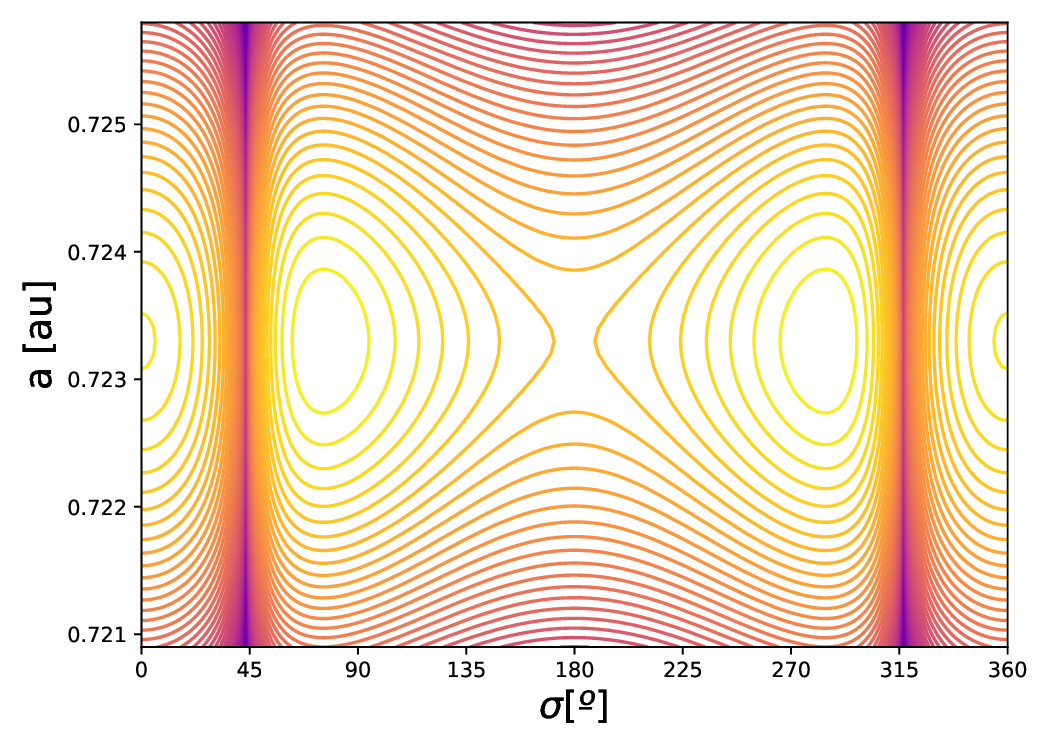}
  \end{minipage}%
  \begin{minipage}[c]{0.45\textwidth}
    \centering \includegraphics[width=3in]{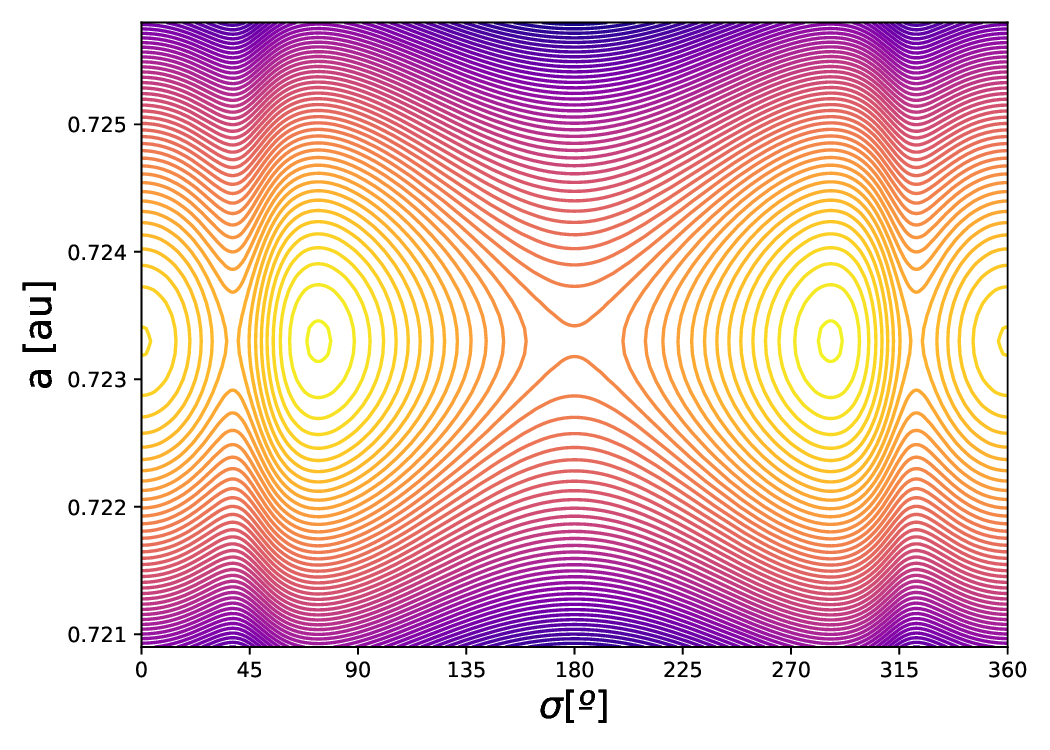}
  \end{minipage}
  \begin{minipage}[c]{0.45\textwidth}
    \centering \includegraphics[width=3in]{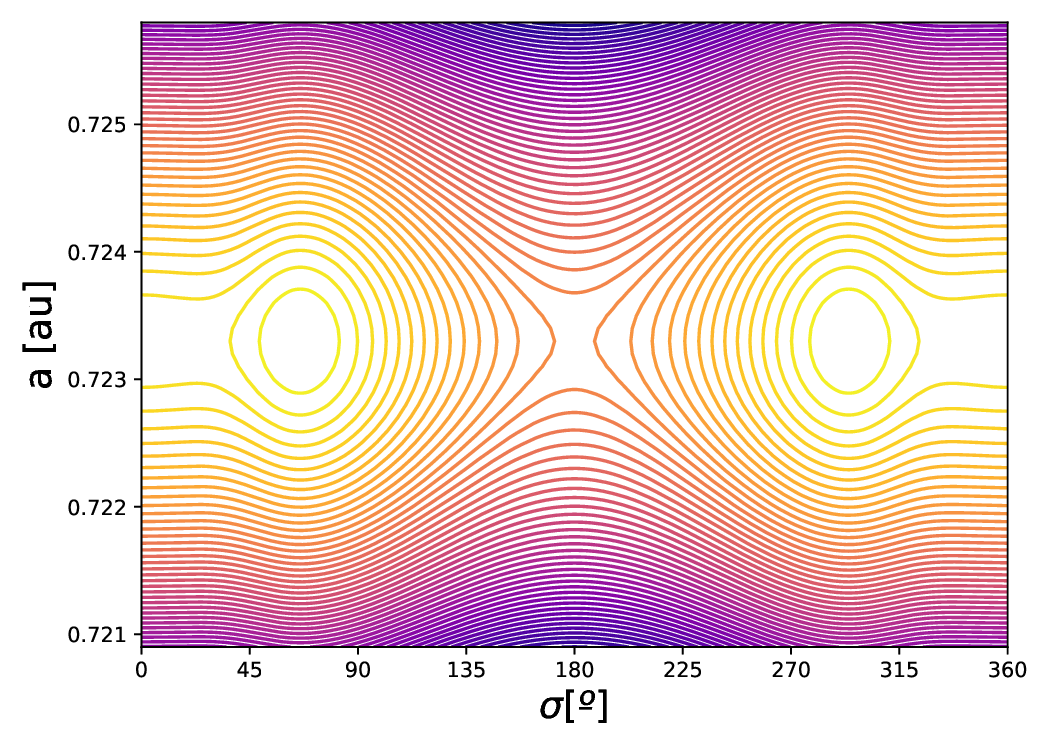}
  \end{minipage}%
  \begin{minipage}[c]{0.45\textwidth}
    \centering \includegraphics[width=3in]{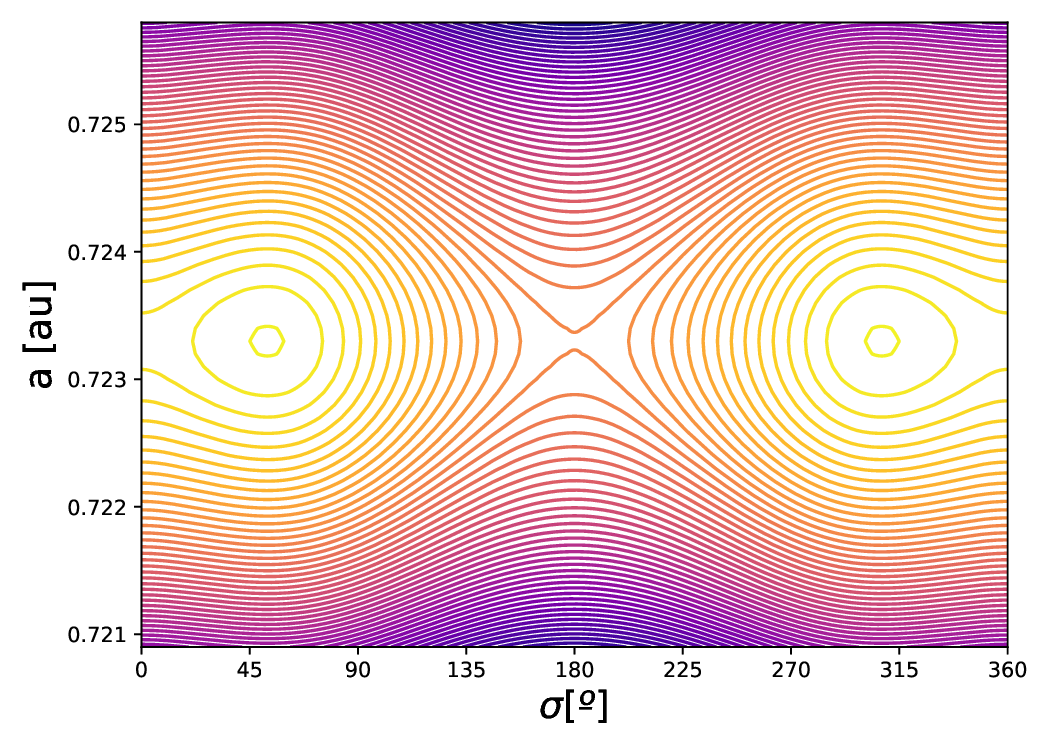}
  \end{minipage}
  \caption{Hamiltonian level for $e = 0.38$ and $inc = 0^{\circ}$ (top left panel), $inc = 15^{\circ}$ (top right panel), $inc = 30^{\circ}$ (bottom left panel), and  $inc = 45^{\circ}$ (bottom right panel).}
  \label{Fig: ham_e_0}
\end{figure*}

What are the dynamics at $e = 0.38$, the limit by which most currently
known Venus' co-orbitals can be observed? Figure~(\ref{Fig: ham_e_0}) displays
the level curves of Hamiltonian~(\ref{eq: Pan_Hamiltonian}) for $e = 0.38$ and the four discussed values of inclination
in the $(\sigma, a)$ plane. At $inc = 0^{\circ}$ we have the three equilibrium
points, distinctly separated by an energy barrier at $\sigma = 45^{\circ}$ and $\sigma = 315^{\circ}$. The situation is similar at $inc = 15^{\circ}$, but compound orbits are now possible across the energy barriers.  At $inc = 30^{\circ}$ the
equilibrium point at $\sigma = 0^{\circ}$ becomes unstable, and that persists at $inc = 45^{\circ}$. Overall, rich and different dynamics can be observed
at the same eccentricity, but for different inclinations.

\begin{figure*}
  \begin{minipage}[c]{0.45\textwidth}
    \centering \includegraphics[width=2.3in]{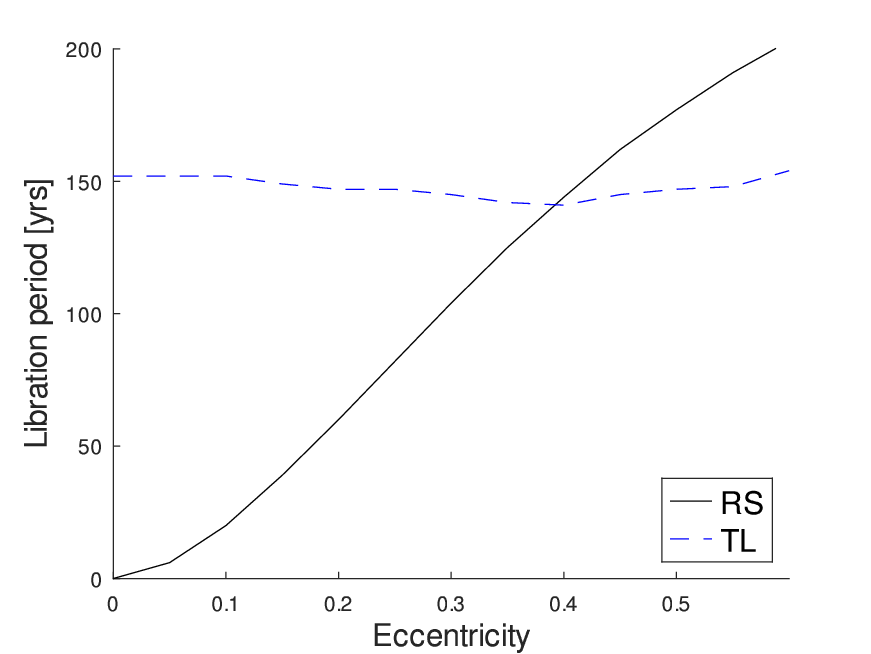 }
  \end{minipage}%
  \begin{minipage}[c]{0.45\textwidth}
    \centering \includegraphics[width=2.3in]{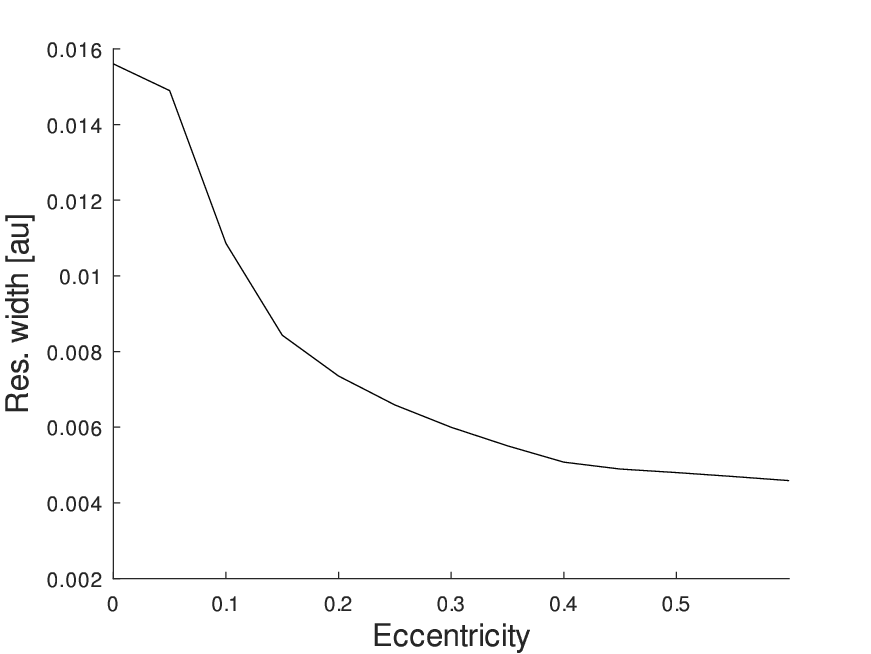}
  \end{minipage}
  \caption{Left panel: libration periods for TL and RS orbits as a function
  of $e$ at $inc = 0^{\circ}$.  Right panel: resonance width as a function of $e$ at the same inclination.}
  \label{Fig: periods_widths}
\end{figure*}

Concerning the libration period for small amplitudes and resonance width,
Figure~(\ref{Fig: periods_widths}) displays how such quantities vary as
a function of $e$ for $inc = 0^{\circ}$.  The libration periods for
TL orbits remain fairly stable, while they increase for larger values
of $e$ for RS orbits.  Please notice that RS orbits are not present at
$e = 0$.  The resonance width is inversely proportional to $e$, with a maximum
value of $e = 0$.  Similar plots were obtained for other values of
$inc$, but are not shown, for brevity.

How reliable is the \citep{pan2024attemptbuilddynamicalcatalog} model for predicting the instantaneous orbital evolution of Venus' co-orbitals?  \citep{Namouni1999a, 2000Icar..144....1C} distinguished between these types of orbits for co-orbitals of terrestrial planets:

\begin{itemize}
\item Retrograde satellites (RS), whose orbit fluctuates around $0^{\circ}$ in the $(\sigma, a)$ plane and stay near to the perturbing planet.
\item Tadpole orbits (TL4 and TL5), characterized by their resemblance to a tadpole in the Sun-planet rotating frame and by their oscillation around the Lagrangian $L_4$ or $L_5$ equilibrium points.
\item Horseshoe orbits (H) appear like a horseshoe loop in the rotating frame because they never close the space at $0^{\circ}$ with respect to the planet. There isn't a presently known Venus co-orbital in this configuration.
\item H-RS compound orbits, which are mergers of RS and H, or H and RS.
\item T-RS compound orbits, resulting from a combination between Tadpoles and RS.
\item Passing orbits (P), or circulating ones (C), for objects not in resonance.
	\end{itemize}

There is also the possibility of transition orbits between any of these configurations.  Figure (1) of \citet{Carruba2025} displays examples of each one of these orbital configurations. Twenty co-orbital asteroids of Venus are currently known (see \citet{Carruba2025} and references therein.

To test the \citep{pan2024attemptbuilddynamicalcatalog} model performance in reproducing the dynamics of known asteroids, we applied this approach using the initial conditions for the 20 currently known co-orbitals of Venus, as reported in Table (1) of \citet{Carruba2025}, as initial conditions for both the semi-analytical model and for numerical simulations with the SWIFT$\_$BS integrator, which is a Burlisch-Stoer integrator of the SWIFT package \citep{2013ascl.soft03001L} with a tolerance factor of $10^{-8}$. Asteroids were integrated over 1000 years under the gravitational influence of all planets.

\begin{figure*}
  \centering
  \begin{minipage}[c]{0.45\textwidth}
    \centering \includegraphics[width=2.3in]{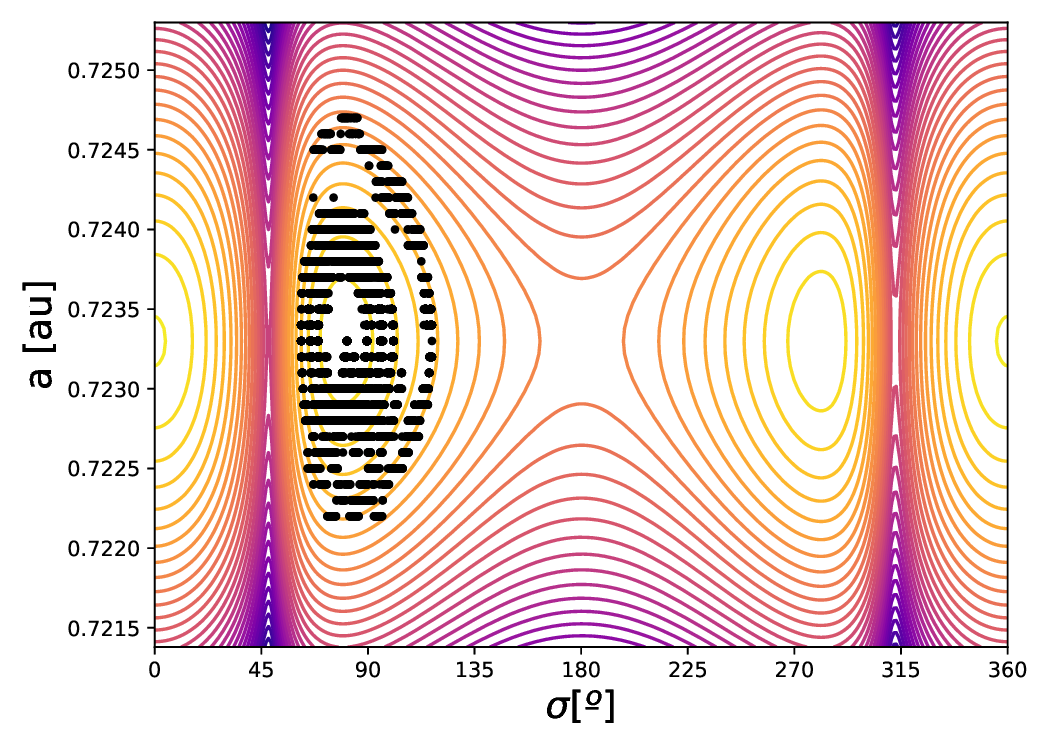}
  \end{minipage}%
  \begin{minipage}[c]{0.45\textwidth}
    \centering \includegraphics[width=2.3in]{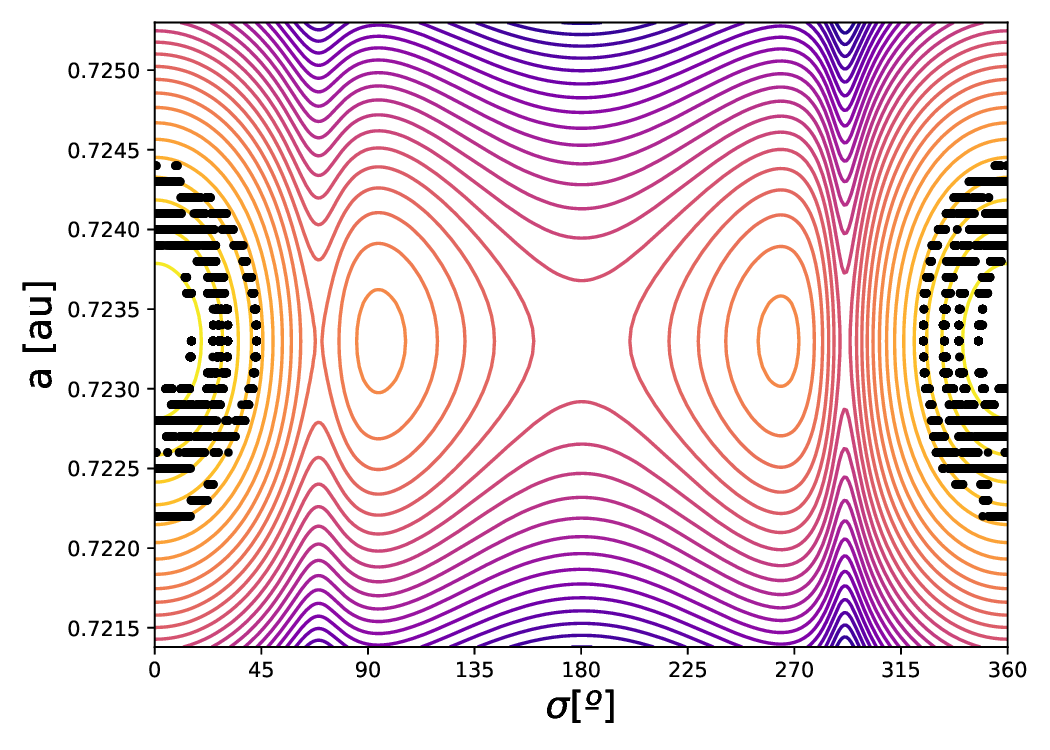}
  \end{minipage}
  \begin{minipage}[c]{0.45\textwidth}
    \centering \includegraphics[width=2.3in]{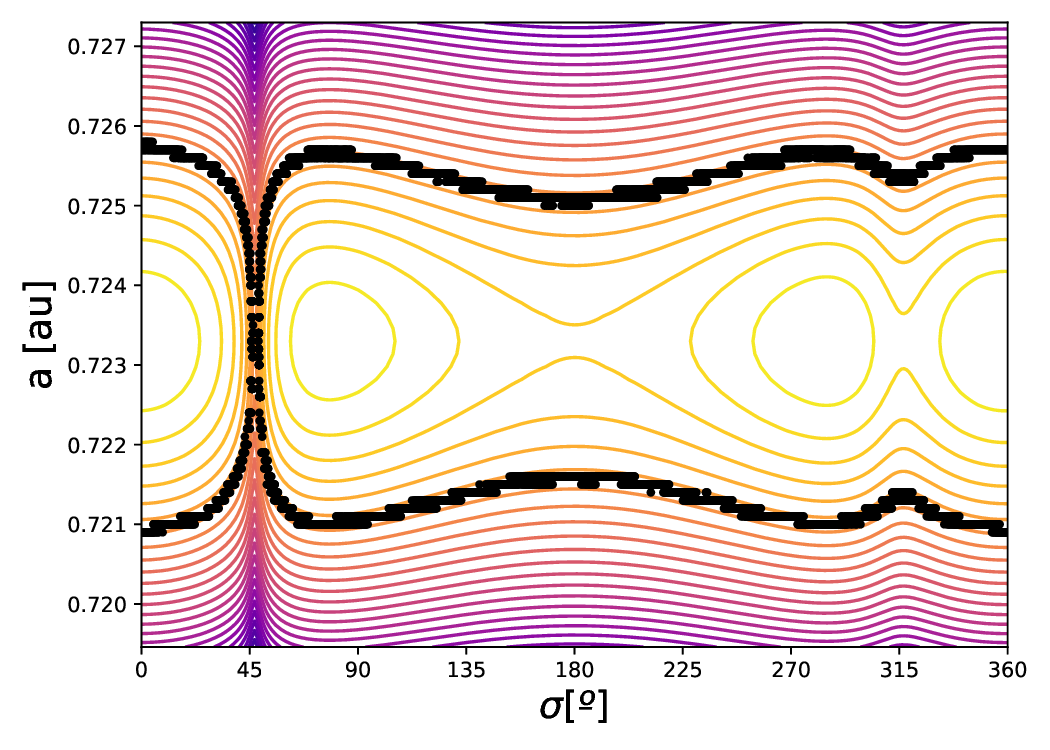}
  \end{minipage}%
  \begin{minipage}[c]{0.45\textwidth}
    \centering \includegraphics[width=2.3in]{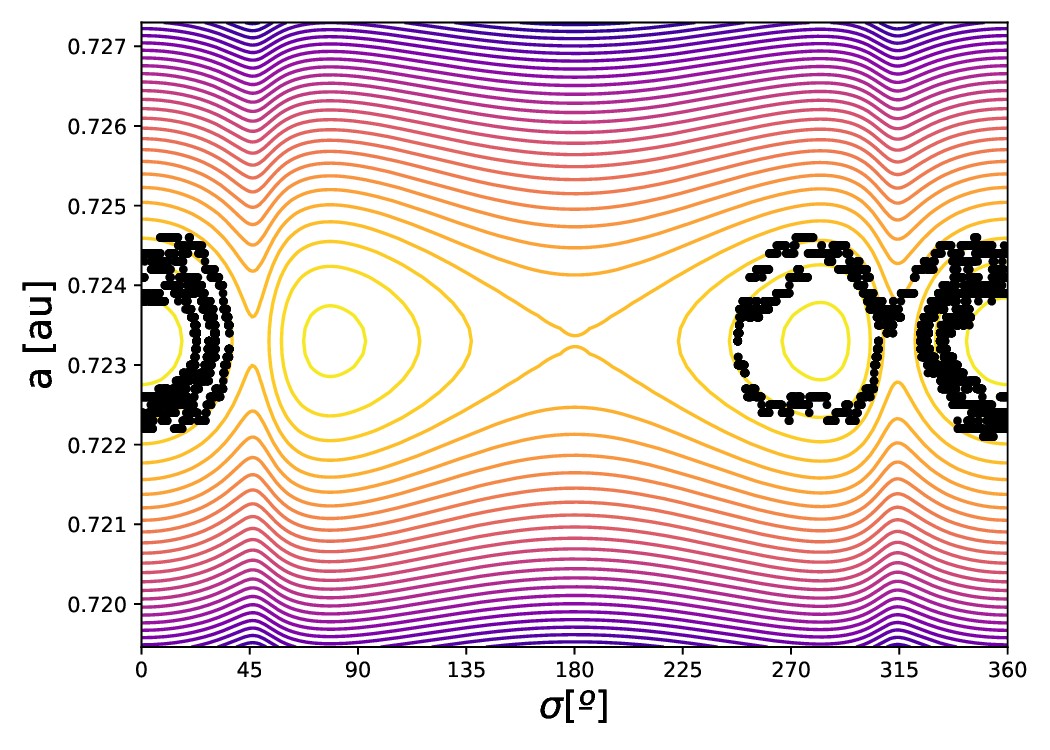}
  \end{minipage}
  \caption{Projections in the $(\sigma, a)$ plane of the Hamiltonian levels and
    the output of numerical simulations (black dots) for an asteroid in a TL4
    configuration (top left panel), in an RS orbit (top right panel), in an
    HRS configuration (bottom left panel) and in a TRS orbit (bottom right
    panel).}
  \label{Fig: pan_real_asteroids}
\end{figure*}

Figure~(\ref{Fig: pan_real_asteroids}) displays the orbital evolution of four asteroids, one inside a TL4 configuration (2020 BT2), one inside an RS orbit (2023 BB1), another in an Horseshoe-Retrograde Satellite (HRS) configuration (2024 AF6), and one in a Tadpole-Retrograde Satellite (TRS) orbit (524522). While the semi-analytical model is instantaneous, and the numerical simulations covered 1000 years, the dynamic is adiabatic over this period, i.e. changes occur slowly enough that the system can adjust without losing energy \citep{2022CeMDA.134...23F}, and the outcome of the numerical simulations tend to shift between near energy levels. Moreover, a very small discrepancy could occur due to the perturbations included in the simulations.  Based on this analysis, it seems reasonable to use the \citep{pan2024attemptbuilddynamicalcatalog} approach to create initial conditions for fictitious, yet unobserved Venus'co-orbitals.  The method used for this purpose will be discussed in the next section.

\section{Close encounters with Earth}
\label{sec: close_enc}

The main goal of this research is to assess the collisional hazard that the undetected population of Venus co-orbital may pose to Earth.  As discussed in \citet{Carruba2025}, the instantaneous present configuration of co-orbital asteroids will change in the future, with objects alternating between different
orbital types in a timescale of the order of 1000 years.  The co-orbital asteroids of Venus are highly chaotic, with Lyapunov times of the order of $\simeq$ 150 years \citep{Tancredi1998}. As a consequence, the study of a single orbit yields little information for times longer than $\simeq~150$~years, but statistical studies of clones of asteroids may provide insights on the long-term dynamics \citep{2000Icar..144....1C}.

\begin{figure}
  \centering
  \includegraphics[width=3in]{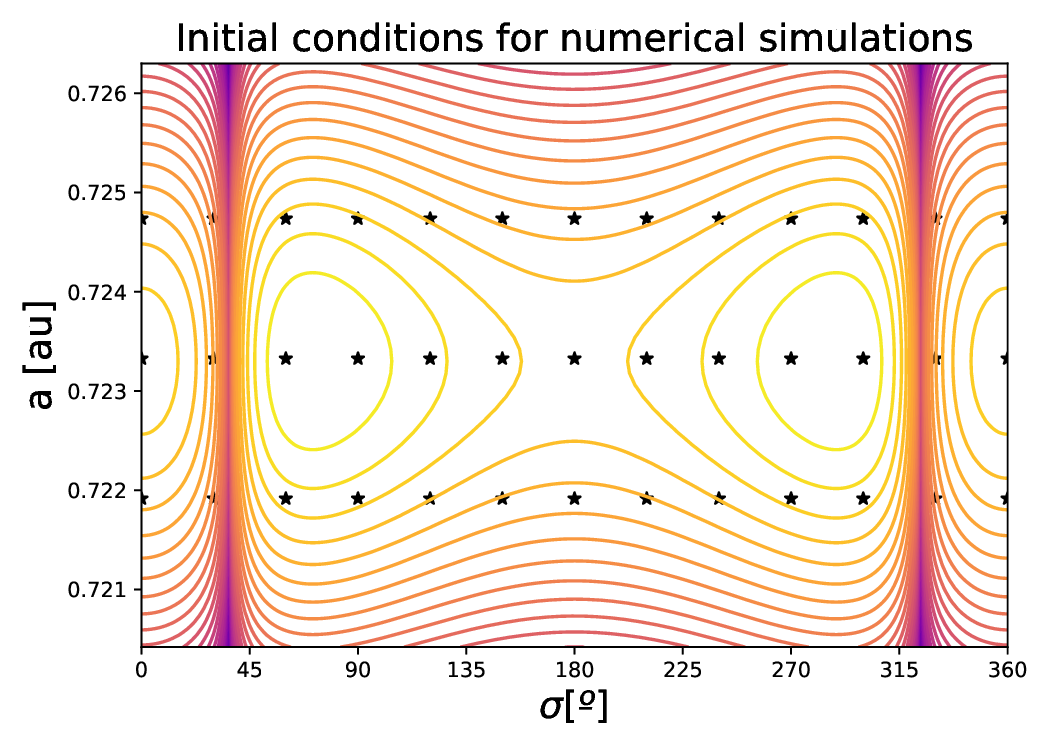}
  \caption{Initial conditions for 36 clones co-orbital asteroids with $e=0.32$ and $i=0^{\circ}$ in the $(\sigma, a)$ plane, shown as black asteroids, superimposed to phase space Hamiltonian level curves obtained using the \citep{pan2024attemptbuilddynamicalcatalog} approach for Venus.}
  \label{Fig: venus_ic}
\end{figure}

To investigate the long-term collisional danger posed by Venus co-orbital asteroids, we therefore used
this approach:  we created a grid of 19 by 5 intervals in the $(e, inc)$ domain, with $inc$ the inclination referred to the instantaneous orbital plane of Venus, using a step of 0.02 in $e$ and $15^{\circ}$ in $inc$, starting at 0.24 in $e$ and $0^{\circ}$ in $inc$.  For each set of $(e, inc)$ values, we created 36 clones in the $(a, \sigma)$ domain, with 12 equally spaced values in $\sigma$, with a step of $30^{\circ}$ and three intervals in $a$, equal to $a_v-da/4, a_v, a_v+da/4$, where $a_v$ is the current value of Venus semi-major axis, and $da$ is the resonance width, as predicted by the \citep{pan2024attemptbuilddynamicalcatalog} approach.  We assume that the mean anomaly $M$, the argument of pericenter $\omega$, and the longitude of the node $\Omega$ are initially equal to zero. Since the test particles all undergo several co-orbital cycles during the integrations, initial values of $M$, $\omega$, and $\Omega$ matter little to track close encounters with Earth, as also confirmed by numerical investigations with different initial values of these angles.  An example of initial conditions for $e=0.32$ and $inc=0^{\circ}$ is displayed in Figure~(\ref{Fig: venus_ic}).  We found that this test particle coverage is large enough to account for the different types of co-orbital behavior observed in a cycle and sufficient to track the effects of close encounters with Earth.

We then integrated the clones over the influence of the eight planets with the SWIFT$\_$BS integrator over 36000 years, which is three times the length of a co-orbital cycle for a Venus'co-orbital, as found in \citet{Carruba2025}.  We neglect non-gravitational forces like the Yarkovsky force because over such timescales, their effects should be limited \citep{Carruba2025} for objects larger than 50 m. Finally, we monitored if an asteroid had a close encounter with Earth
by checking if its orbital distance to the planet was less than the Hill's
sphere radius, and tracked what was the minimum orbit intersection distance (MOID) during the simulation. Interested readers can find more details about this procedure in \citet{Carruba2025}.

\begin{figure*}
  \centering
  \includegraphics[width=4.6in]{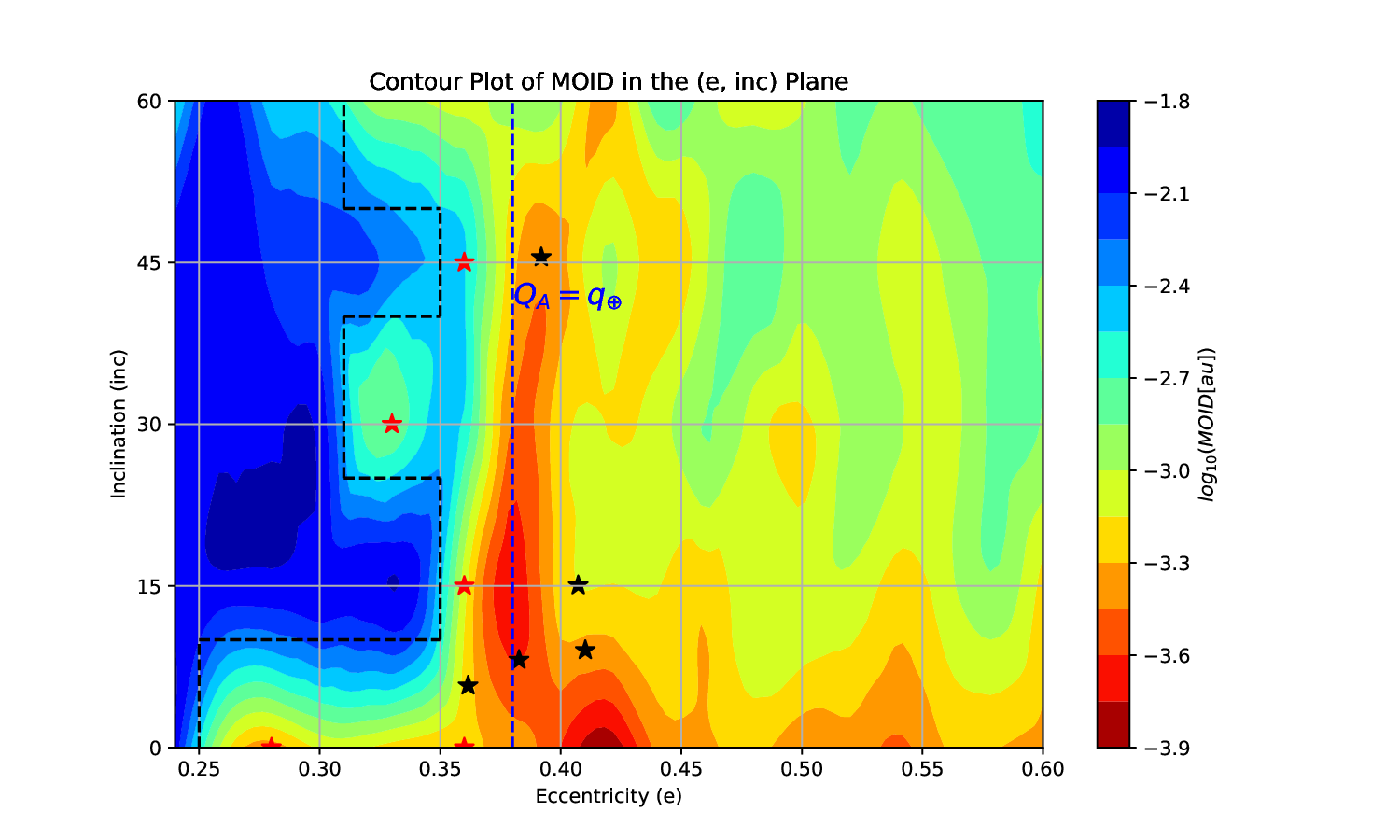}
  \caption{Contour plot of the MOID with Earth as a function of the initial $(e, inc)$ values.  The vertical line displays $e=0.38$, for which the apocenter of the Venus co-orbital asteroid is equal to Earth's pericenter.  The black stars display the location of the real co-orbitals of Venus with a MOID with Earth of 0.0005 au or less.  The red stars show the orbital location of five test particles at $e < 0.38$ that experienced low MOIDs in our simulations. The dashed black line shows the approximate boundary of the low-MOID region.}
  \label{Fig: venus_enc_map}
\end{figure*}

Figure~(\ref{Fig: venus_enc_map}) displays a contour plot of the minimum MOID with Earth in the initial $(e,inc)$ plane. As expected, most close encounters occur for $Q_A = q_{\oplus}$, i.e. $e=0.38$. However, there exists a portion of the phase space at lower eccentricities where possible co-orbital asteroids can experience a large number of close encounters and, possibly, collisions with Earth.  The width in $e$ of this space is greater at lower inclination values, with a minimum eccentricity of 0.26 observed for $inc = 0^{\circ}$. These findings are consistent with what was detected by \cite{de2023orbital} and \cite{araujo2018rings}, which showed that orbits with lower inclinations are more susceptible to close planetary encounters. For inclinations higher than $10^{\circ}$, the shape of the region with a low MOID with Earth remains essentially constant and limited to $0.35 < e < 0.38$ up to $inc = 25^{\circ}$.  Then, a region with slightly lower values of MOID is observed between $inc = 25^{\circ}$ and $inc = 40^{\circ}$ for $e$ in a range of 0.31 to 0.38.  This region is not observed in contour plots of the number of encounters, and may be a numerical fluke.  Nevertheless, it is outlined in Figure~(\ref{Fig: venus_enc_map}) through black dashed lines. Between $inc = 40^{\circ}$ and $inc = 50^{\circ}$ the limits of the low-MOID region go back to a range $ 0.35 < e < 0.38$.  Finally, for $inc > 50^{\circ}$, a value larger than that of the most highly inclined known Venus co-orbital asteroid (2020 CL1, with $inc= 45.49^{\circ}$), the low-MOID region appears to grow bigger once again. 
Five of the $PHAs$ identified in \citet{Carruba2025} are all in the region
of low MOID displayed in Figure~(\ref{Fig: venus_enc_map}), and this confirms its potential as a basis for predicting the behavior of other yet undetected $PHAs$.

To check for the observability of potentially undetected Venus's co-orbitals, we investigated the behavior of particles in the black dashed-line region
  in Figure~(\ref{Fig: venus_enc_map}), and we
also selected five test particles that had $e< 0.38$ and experienced i) the
largest number of close encounters with Earth and ii) the encounters at a minimum
relative distance. The five test particles have $e = 0.28$ and $inc =0^{\circ}$ (fic1), $e = 0.36$ and $inc =0^{\circ}$ (fic2), $e = 0.36$ and $inc =15^{\circ}$ (fic3), $e = 0.33$ and $inc =30^{\circ}$ (fic4), and $e = 0.36$ and $inc =45^{\circ}$ (fic5).  They are displayed as black stars in Figure~(\ref{Fig: venus_enc_map}).

\section{Observability from Earth}
\label{sec: observ_earth}

Recent surveys have used the Blanco 4-meter telescope with the Dark Energy Camera 
\citep{Sheppard_2022_AJ_164_168} and the Palomar 48-inch telescope 
(P48)/Zwicky Transient Facility (ZTF) \citep{Bolin-et-al-2025} to look for objects inside Earth's and Venus' orbits. 
The first survey discovered two Atira asteroids and an Apollo, inclusive of 2021 PH27, which suffers the highest 
relativistic effects from the Sun. The second survey discovered four Atira objects and the first asteroid with an orbit completely inside that of Venus, (594913) ‘Ayl\'{o}’chaxnim.

The "low-SE twilight survey" is a special observation program proposed by the Rubin Observatory user community. The goal will be to take short exposures closer to the Sun to look
for small celestial bodies in an orbital space that the rest of the Rubin Observatory cannot detect, and this could potentially 
significantly increase our knowledge of asteroids near Venus \citep{Schwamb2023}.

{To assess the feasibility of detecting the unobserved population of Venus co-orbital 
asteroids, we also analyzed their observability from the Vera C. Rubin Observatory's 
regular observing program, located at Cerro Pachón, Chile (latitude $30.24^\circ$~S, 
longitude $70.75^\circ$~W, elevation 2647 m).}

{For this analysis, we considered both the known Venus co-orbital 2020 CL1 and five 
fictitious objects representing potential undetected populations. Table \ref{T-obs} 
presents their orbital and physical parameters, along with key observability metrics 
considering only periods when targets are above $20^\circ$ elevation. For 2020~CL1, 
all orbital elements were obtained from JPL Horizons, while for the fictitious 
objects, the angular elements (argument of pericenter, longitude of node, and mean 
anomaly) were set to zero. The test particles span orbital inclinations from 
$0^\circ$ to $45^\circ$, while maintaining semi-major axes near Venus's 
orbit ($\sim$0.72 au) and eccentricities below the Earth-crossing threshold of 
0.38. We computed their positions and apparent magnitudes from January 2020 to 
December 2035 using JPL Horizons system (using DE441 planetary ephemeris and 
small-body perturber SB441-N16), excluding daylight hours. The apparent magnitudes 
were derived using the standard IAU H-G system. For the fictitious objects, we 
adopted an absolute magnitude $H=22$, which is the limit for an asteroid to be considered a $PHA$ \citep{PHA_task_force_2000}.}

\begin{table*}[!ht]
\centering
\begin{tabular}{lccccccc}
\hline
Object & $H$ & $a$ (au) & $e$ & $inc$ ($^\circ$) & \multicolumn{3}{c}{At maximum brightness after Jan/2025} \\
\cline{6-8}
& & & & & Elevation ($^\circ$) & Magnitude & Elongation ($^\circ$) \\
\hline
2020 CL1 & 19.95 & 0.724 & 0.39 & 45.49 & 20.19 & 17.65 & 98.02 \\
fic1 & 22.00 & 0.723 & 0.28 & 0.00 & 25.81 & 20.10 & 71.43 \\
fic2 & 22.00 & 0.725 & 0.36 & 0.00 & 29.02 & 21.20 & 66.61 \\
fic3 & 22.00 & 0.724 & 0.36 & 15.00 & 21.73 & 20.15 & 76.75 \\
fic4 & 22.00 & 0.724 & 0.33 & 30.00 & 20.83 & 20.53 & 80.15 \\
fic5 & 22.00 & 0.724 & 0.36 & 45.00 & 25.64 & 20.27 & 79.60 \\
\hline
\end{tabular}
\caption{Orbital and observability parameters for real (2020 CL1) and fictitious 
Venus co-orbitals. The columns give the absolute magnitude ($H$), semi-major axis 
($a$), eccentricity ($e$), and inclination ($i$). The last three columns show each 
object's parameters at its maximum brightness when above $20^\circ$ elevation after 
January 2025.}
\label{T-obs}
\end{table*}

\begin{figure}[!ht] 
\centering \includegraphics[width=0.95\columnwidth]{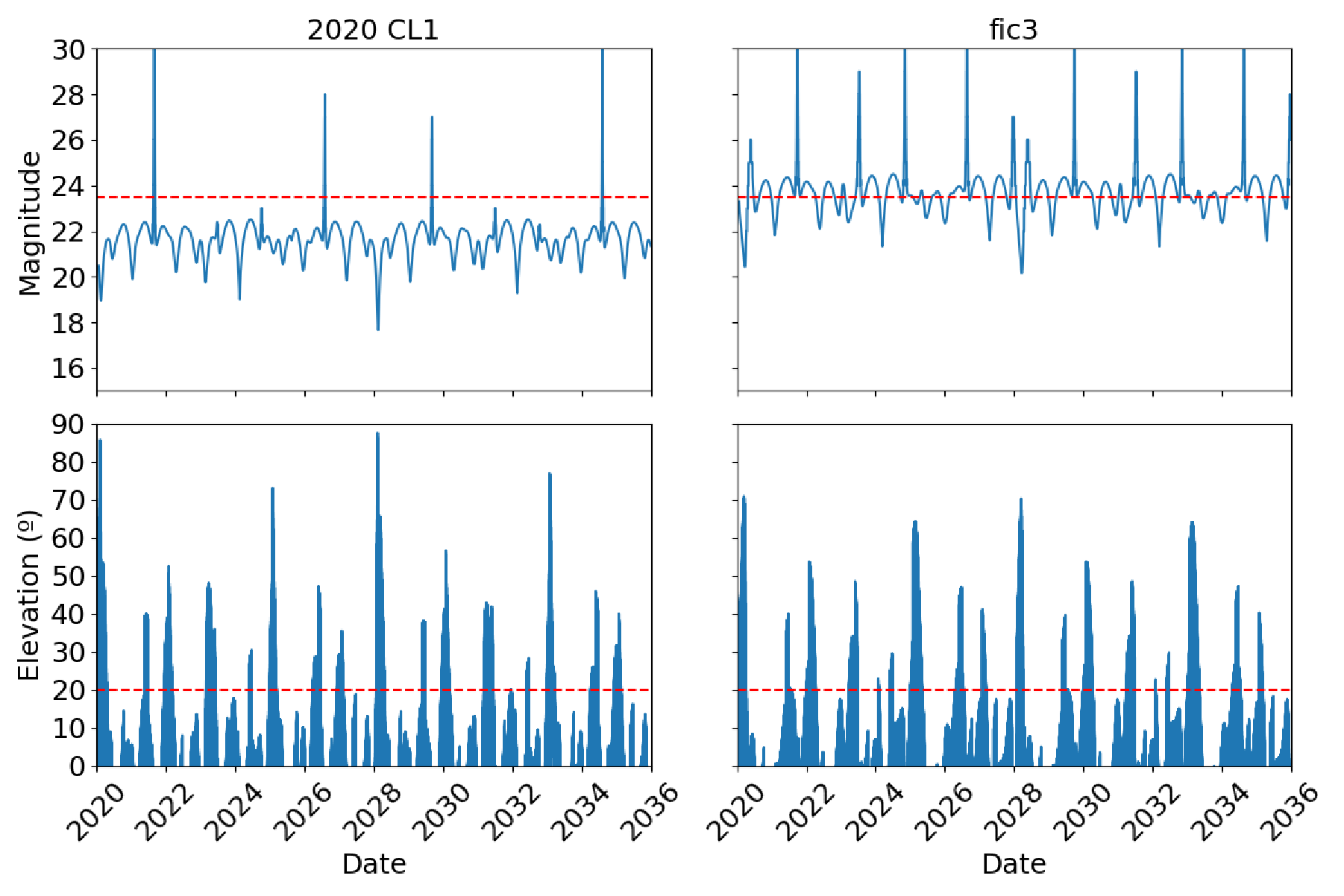} 
\caption{Observability conditions for Venus'co-orbitals from the Rubin Observatory site over 
2020-2036. The left panels show the known co-orbital 2020~CL1, while the right panels 
show the fictitious object fic3. Upper panels display the apparent magnitude 
variations, with the horizontal dashed line indicating Rubin Observatory single-visit detection limit of 23.5. Lower panels show elevation above the horizon, where the 
dashed line marks the $20^\circ$ minimum elevation required for effective 
observations.}
\label{Fig: mosaic}
\end{figure}

{To exemplify the observational constraints, Fig. (\ref{Fig: mosaic}) shows apparent 
magnitude and elevation plots for 2020 CL1 and one test particle, fic3, as viewed 
from the Rubin Observatory site. Both objects present several observational windows throughout the studied period, when they exceed the minimum elevation threshold of 
$20^\circ$, as indicated by the horizontal dashed line in the bottom panels. 
During these windows, which typically last for a few weeks, their apparent 
magnitudes remain below or close to the Rubin Observatory single-visit detection limit of 23.5 \citep{ivezic19}.}

After January 2025 and considering observability windows with elevation above 
$20^\circ$, the apparent magnitudes of our test particles reach their minimum values between 20 and 21. At these maximum brightness configurations, the elevations range from $20.8^\circ$ to $25.8^\circ$, with solar elongations between $69.2^\circ$ and $80.2^\circ$. The variations follow a similar pattern to that observed for 2020 CL1, although this real object, with $H=19.95$, reaches magnitude 17.65 at $20.2^\circ$ elevation and significantly larger elongation of $98.0^\circ$.

An interesting pattern emerges when examining the relationship between orbital 
inclination and observability conditions. While objects of different inclinations  achieve similar elevations ($\sim 20^\circ$-$26^\circ$) at maximum brightness, 
their solar elongations show that the low-inclination objects ($inc \sim 
0^\circ$) are observed at slightly smaller elongations ($\sim 70^\circ$), while objects with higher inclinations ($inc > 15^\circ$) cluster around $80^\circ$ elongation. 

The elongation values also reveal an important distinction between natural and 
fictitious objects. Although all fictitious objects with $inc > 15^\circ$ maintain similar elongations at maximum brightness (clustered within $\sim1^\circ$ of each other), 2020 CL1's significantly larger elongation ($98.02^\circ$) suggests that there may be a possible current observational bias toward discovering objects at larger solar elongations, since it is hard to perform observations at low solar elongation, which can be done only for a few hours at dawn/sunset. This may be a key factor in explaining the limited number of known Venus co-orbitals.

The analysis of all fictitious objects reveals similar general patterns in their observability. Objects with similar inclinations (fic1-fic2 and fic4-fic5) show comparable temporal behavior in both magnitude variations and elevation patterns. The low-inclination objects present more frequent but slightly shorter observable windows compared to their high-inclination counterparts. All fictitious objects exhibit periodic brightness variations driven by their orbital geometry relative to Earth, with maximum brightness generally occurring near their closest approaches to Earth while remaining above the minimum elevation threshold.

Looking at the temporal behavior shown in Fig. \ref{Fig: mosaic}, both 2020~CL1 and fic3 present periodic patterns in their apparent magnitudes and observable windows, though with distinct characteristics. While 2020 CL1 shows relatively uniform magnitude variations between 20-22, fic3 exhibits larger amplitude variations with occasional sharp brightness peaks. Despite these differences, both objects remain consistently brighter than Rubin Observatory detection limit of 23.5, suggesting that detection should be feasible even under less-than-optimal conditions.

The combination of elevation constraints and solar elongation limitations restricts observations to specific periods throughout the year. These constraints particularly affect the detection of objects with lower orbital inclinations, which 
must be observed at relatively small solar elongations even during their most favorable configurations.

To assess the observability of Venus co-orbital asteroids systematically, we analyzed a statistical sample of test particles that experienced low MOIDs with Earth. Particles were selected from five inclination values ($0^\circ$, $15^\circ$, $30^\circ$, $45^\circ$, and $60^\circ$), shown as red stars in Fig.\ref{Fig: venus_enc_map}, with various combinations of semi-major axis, eccentricity (ranging from 0.26 to 0.38 in steps of 0.02), and mean anomaly, as delimited by the dashed black line in Figure~(\ref{Fig: venus_enc_map}). For each object, we computed ephemerides from the Vera C. Rubin Observatory site, applying sequential visibility filters: objects above the horizon, with elevation $> 20^\circ$ and apparent magnitude $< 23.5$.

\begin{figure}[!ht] 
\centering \includegraphics[width=0.95\columnwidth]{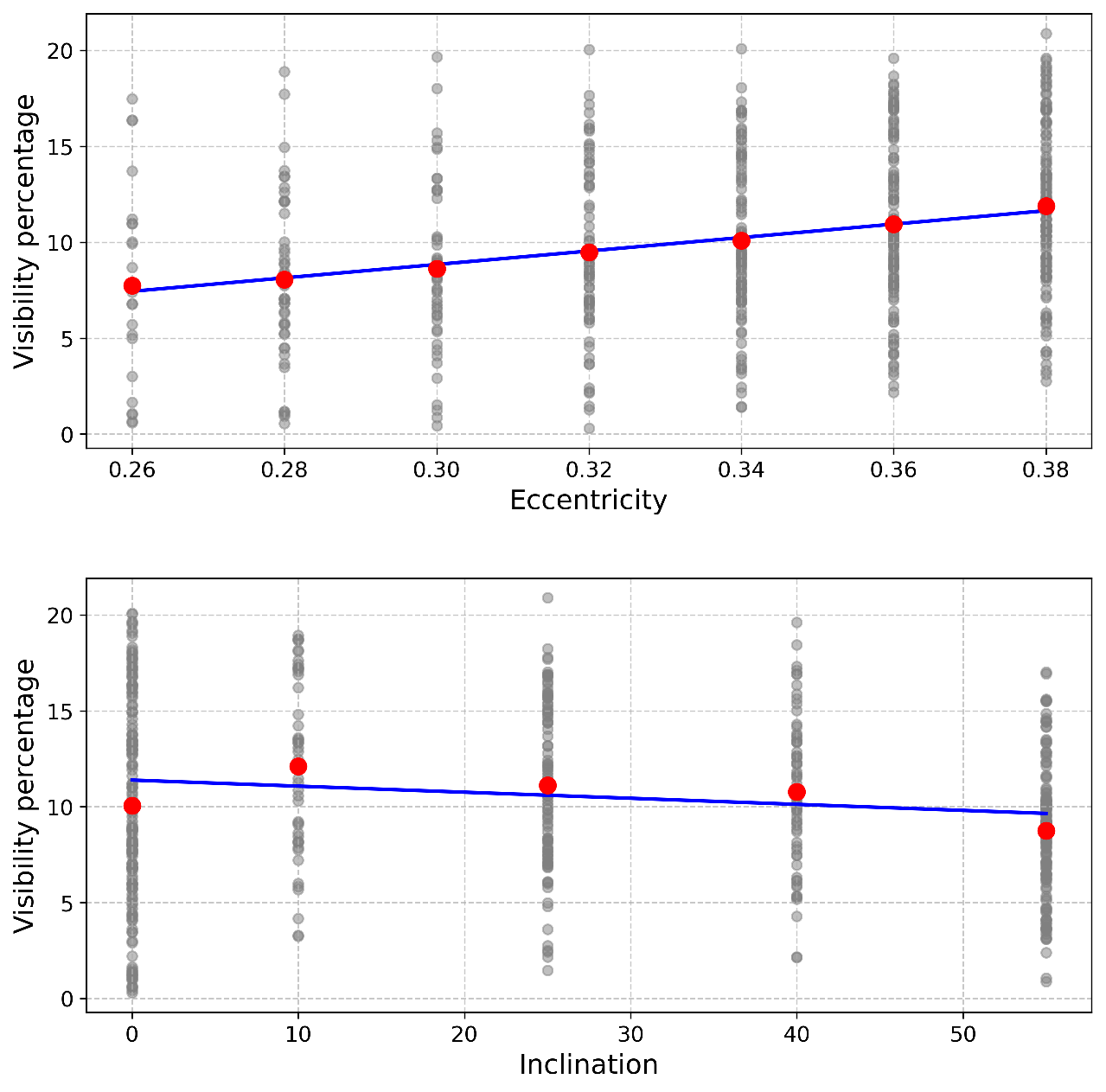} 
\caption{Visibility percentage of Venus co-orbital asteroids as observed from the Rubin Observatory site as a function of orbital parameters. Each gray point represents a simulated object from the statistical sample that experienced low MOIDs with Earth.  The red circles show the average visibility percentage for each eccentricity value (top panel) and inclination value (bottom panel). The blue lines represent linear fits to these averaged values.}
\label{Fig: fit}
\end{figure}

We calculated the "visibility percentage" – defined as the fraction of total ephemeris entries that satisfy all observability criteria – as a function of orbital elements. This metric effectively measures the fraction of time an object would be detectable by the Rubin Observatory during favorable observing windows. For each eccentricity and inclination value, we computed the average visibility percentage across all objects sharing those orbital parameters. We then performed linear fits to these averaged values to identify trends, as shown in Fig.~\ref{Fig: fit}. The analysis reveals a strong positive correlation between eccentricity and visibility percentage, with higher eccentricity objects being observable longer. This confirms the observational bias favoring high-eccentricity Venus co-orbitals. In contrast, no statistically significant trend is observed between inclination and visibility percentage with values consistently around 10\% across all inclination values, indicating that orbital inclination has minimal impact on the detectability of Venus co-orbitals from Earth-based observatories.

This behavior has a geometric explanation: objects with higher eccentricities (until $e = 0.38$) achieve closer approaches to Earth's orbit, resulting in smaller Earth-asteroid distances during favorable alignments. This proximity produces brighter apparent magnitudes and increased detection probability. The inclination independence suggests that the Earth-asteroid distance dominates detectability rather than out-of-ecliptic deviations. 

\section{Observability from Venus' orbit}
\label{sec: observ_venus}

\begin{figure}[!ht]
\centering
\centering \includegraphics[width=0.95\columnwidth]{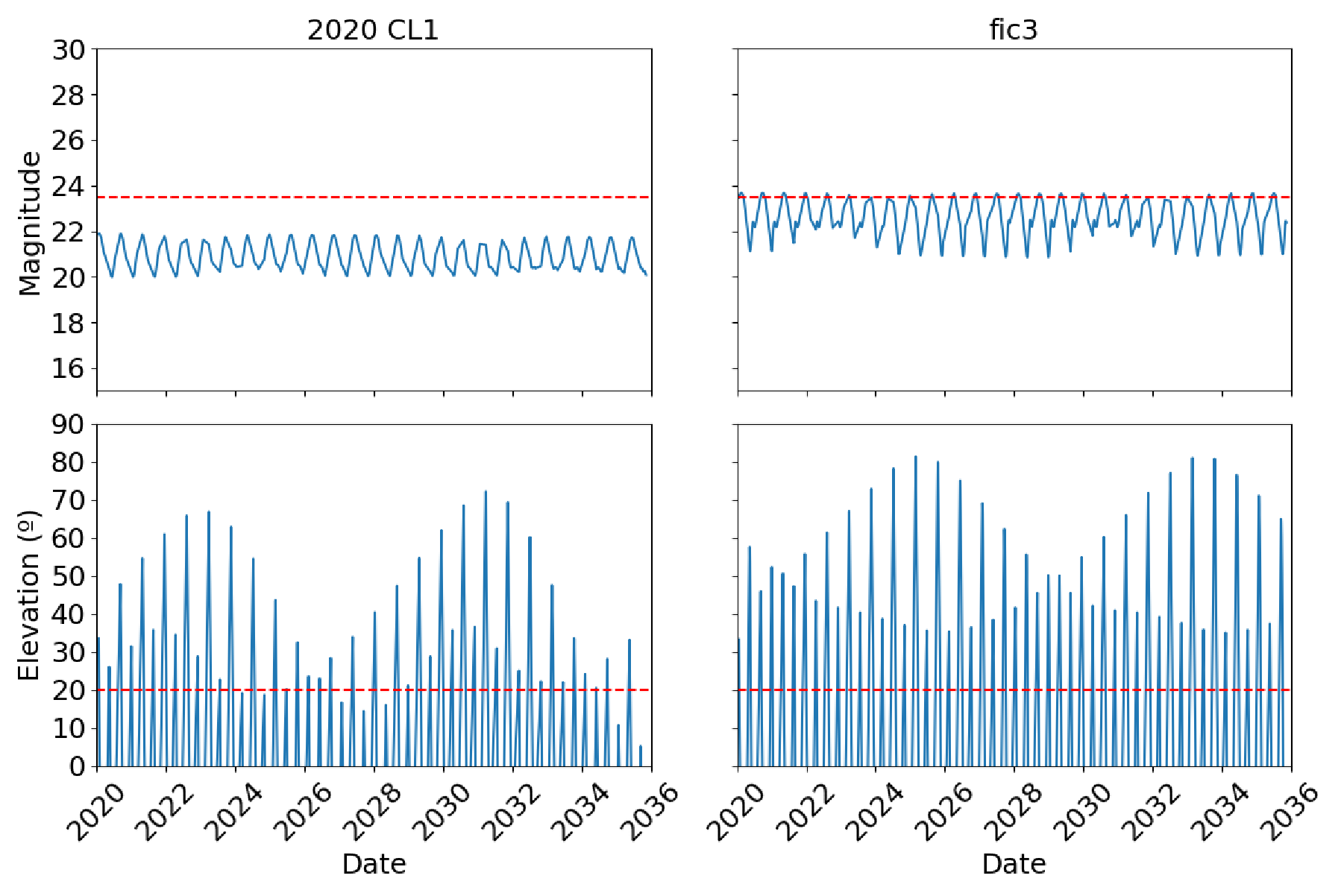}
\caption{Observability conditions for Venus co-orbitals from Venus. The left panels show the known co-orbital 2020 CL1, while the right panels show the fictitious object fic3. Upper panels display apparent magnitude variations, with the 
horizontal dashed line indicating a hypothetical detection limit of 23.5. Lower panels show elevation above the horizon, where the dashed line marks the 20° minimum elevation required for effective observations.}
\label{fig:mosaic_venus}
\end{figure}

As seen in the previous section, detecting inner asteroids from Earth, including Venus' co-orbital asteroids, is challenging. Due to their proximity to the Sun and relatively fast orbital motions, which causes a fast apparent motion in the sky, these objects require advanced observational techniques. However, observations conducted from Venus's orbit, positioned facing away from the Sun, may enhance the detection of these bodies.  We performed a similar analysis as in the previous section, but from a hypothetical observer on Venus at the equator in the prime meridian to check the observability conditions. Of course, because of the thick and corrosive atmosphere of Venus, no telescope could be viable or survive long from the surface \citep{Lukco2018}.  Nevertheless, estimates of visibility obtained from such location could provide insights into what a telescope in orbit around the planet or in the near $L_1$ and $L_2$ Lagrangian point could see. Figure~(\ref{fig:mosaic_venus}) displays the apparent magnitude and elevation of 2020 CL1 and fic3, as observed from Venus. These are the same objects shown in Figure~(\ref{Fig: mosaic}), but now as seen from Venus\footnote{The following analysis is based on the observability of asteroids in the visible spectrum.  Some space telescopes, like NEOWISE \citep{2014ApJ...792...30M} or NEO Surveyor \citep{Hoffman_NEOSurveyor_Overview_2021} operate in the infrared band and are subject to different constraints than the visual magnitude limits discussed here. Nevertheless, for simplicity, here we limit our analysis to the visible part of the spectrum.}.

The apparent magnitude variations seen from Venus exhibit more regular patterns compared to Earth-based observations, with values typically ranging between 20-22 for 2020 CL1 and 18-24 magnitudes for fic3. This stability in brightness is particularly notable for 2020 CL1, which shows much more consistent magnitude variations when observed from Venus compared to Earth-based observations. The fictitious object fic3 displays greater magnitude variations, with periodic brightening events reaching almost magnitude 18.

The elevation angles above 20° (indicated by the horizontal dashed line) occur 
with regular patterns when observed from Venus. 2020 CL1 exhibits two main 
periods of favorable observing windows, peaking at elevations of approximately 
70°, while fic3 shows more frequent high-elevation opportunities, with maximum 
values exceeding 80°. This significant difference in maximum elevation between 
the objects is likely related to their different orbital inclinations, with fic3's $15^\circ$ inclination providing more favorable geometry for a Venus-based 
observer.

Our analysis of all five fictitious objects and 2020 CL1 shows that objects 
with higher inclinations generally achieve better observability conditions from Venus, with more extended periods above the minimum elevation threshold. Across all objects, the apparent magnitudes remain consistently within the detection range of modern telescopes, with significantly less variation than when observed from Earth.

These characteristics suggest that a Venus-based observer would have more favorable and consistent opportunities for detecting and tracking co-orbital objects, though still subject to fundamental visibility constraints due to solar elongation limits. Some missions were proposed, considering this advantage. They can be divided into two broad categories, one associated with a single spacecraft, usually orbiting around a Sun-Earth or Sun-Venus $L_1$ or $L_2$ Halo orbit, and another associated with a constellation design, where several telescopes look for PHAs from Venus-like orbits.

One of the first work on the first type of mission was proposed by \cite{morrison2009physical}, which calculated the transfer of a spacecraft from Earth to reach a Sun-Venus $L_2$ Halo orbit. Although this work focused on orbital maneuvers, the researchers pointed out that this type of orbit could be advantageous for observing objects interior to the Earth's orbit (IEOs or Atira).

The NEO Surveyor \citep{Hoffman_NEOSurveyor_Overview_2021} mission is a targeted mission within NASA's Planetary Defense Coordination Office (PDCO). The project was approved to enter the preliminary design phase (Phase B) in 2021, following an extensive Concept Development Phase (Phase A). NEO Surveyor will orbit in a Halo orbit around the Sun-Earth Lagrange point $L_1$, and it will possibly launched after 2027. It may observe asteroids at a solar elongation as low as $45^{\circ}$, which should be enough to map most of the potentially dangerous undetected Venus' co-orbital asteroids.

A limitation of a single telescope mission is that sky coverage may be limited, and NEAs and PHAs may be observable for limited times and geometries, so limiting the accuracy of the possible retrieved orbit. According to \cite{dunham2010using}, one of the first registered proposals for a mission to Venus aimed to seek potentially hazardous inner asteroids was the Shield proposal, submitted by the Johns Hopkins University’s Applied Physics Laboratory to the NASA Institute for Advanced Concepts \citep{morrison1992spaceguard}. They proposed deploying three spacecraft into orbits close to Venus, with aphelion around 0.8 astronomical units and facing away from the Sun.

A recent study investigating this concept was that of \cite{shirobokov2020design}.  They presented the design of a space mission to seek potentially hazardous asteroids within the inner solar system through a space telescope placed around the Sun-Venus $L_2$ libration point. They found Halo and Lissajous orbits and discussed the station-keeping cost to stabilize such orbits. Similarly, \cite{wilmer2024sun} presented $37$ periodic orbits in the Sun-Venus CR3TB (Circular Restricted Three-Body Problem) that could favor observing inner bodies. 

Finally, \citet{zhou2022near} proposed the mission CROWN, which consists of a constellation of seven objects to be launched after 2031. One of these is a mothership responsible for deployment and maneuverability. At the same time, the other six telescopes, referred to as daughters, are positioned away from the Sun and dedicated to observing inner asteroids from Venus-like orbits. This configuration allows better coverage over the space within the Earth’s orbit, and for better and longer observation arcs.  An analysis of simulated asteroids showed that more than 94.55\% of the targets should be visible for more than 100 days, with only 0.48\% having an observation arc shorter than 20 days, which is unsuitable for orbit determination.  More than 90\% of the simulated objects were visible from at least two telescopes at the same time, allowing for better orbit determination.

\section{Conclusions}
\label{sec: concl}

In this work, we investigated the collisional hazard that an undetected population of Venus'co-orbital asteroids at low eccentricities may pose to Earth.  For this purpose, we:

\begin{enumerate}
\item Revised results from recent modern models for producing NEAs, such as
  NEOMOD3 \citep{2024Icar..41716110N}. The current lack of observations of
  low-$e$ Venus'co-orbitals is unlikely to be caused by dynamic effects, and it is
  most probably to be the product of observational biases.  A large population
  of low-$e$ Venus'co-orbitals could exist but have not been detected.

\item Used the semi-analytical approach of \citet{pan2024attemptbuilddynamicalcatalog} to revise the dynamics of Venus's Trojans near the $Q_A = q_{\oplus}$
  limit at $e = 0.38$.  The \citet{pan2024attemptbuilddynamicalcatalog} model 
  can successfully reproduce the short-term evolution of known Venus Trojans
  in orbital configurations such as TL4, RS, HRS, and TRS (see Figure~(\ref{Fig: pan_real_asteroids})).

\item Used the \citet{pan2024attemptbuilddynamicalcatalog} model to set up initial conditions for a grid of intervals in the $(e, inc)$ domain, to simulate close encounters with Earth over a time-scale of 36000 years, i.e. three times the average length of a co-orbital cycle \citep{Carruba2025}. The majority of near contacts, as anticipated, happen for $Q_A = q_{\oplus}$, i.e. $e=0.38$.  However, there is a sizable population of potential co-orbitals at lower eccentricity capable of numerous close encounters—and perhaps collisions—with Earth. The width in $e$ of this population is larger at lower values of inclinations.

\item {We discussed the feasibility of detecting unobserved Venus co-orbital asteroids from Earth, focusing on the Vera C. Rubin Observatory's regular observation program. The analysis considers a known co-orbital asteroid, 2020 CL1, and five fictional objects, examining their orbital, physical parameters, and observability metrics, including elevation and apparent magnitude. The results indicate that, although observation windows exist, detections are limited by solar elevation and elongation constraints, especially affecting objects with low orbital inclinations. The study also suggests a possible observational bias toward discovering objects at greater orbital eccentricities, since these asteroids are more likely to approach Earth and appear brighter.}

\item Revised current proposals for space missions to observe objects interior to the Earth's orbit.  Space telescopes orbiting the Sun-Venus $L_2$ libration point, or constellations of space telescopes as proposed in the CROWN mission \citep{zhou2022near}, could potentially be very successful in detecting yet undiscovered Venus'co-orbital asteroids. 

\end{enumerate}

The recent discovery of the $PHA$ 2024 YR4, an asteroid that reached the highest ever impact probability for an asteroid larger than 20 meters and is a potential ``city-killer'' (https://www.minorplanetcenter.net/mpec/K24/K24YE0.html), showed that there are still undetected asteroids in the inner Solar System with the potential to cause significant damage to Earth.  Among these, low-$e$ Venus'co-orbitals pose a unique challenge, because of the difficulties in detecting and following such objects from Earth.  While surveys like those from the Rubin Observatory (see Sect.~(\ref{sec: observ_earth})) may be able to detect some of these asteroids in the near future, we believe that only a dedicated observational campaign from a space-based missions near Venus, as discussed in Sect.~(\ref{sec: observ_venus}), could potentially map and discover all the still ``invisible" PHA among Venus'co-orbital asteroids.

\section*{Acknowledgments}

We are grateful to an anonymous reviewer for comments and suggestions that helped increase the quality of this work. VC and OCW acknowledge the support of the Brazilian National Research Council (CNPq, grants 304168/2021-1 and 316991/2023-6). S.D.R is partially supported by G.N.F.M. group of INdAM. GC is grateful to the S\~{a}o Paulo Research Foundation (FAPESP, grant 2021/08274-9). 

\section{Data availability}

Data produced in this work will be made available upon reasonable request.

\section{Code availability}

Codes will be made available upon reasonable request.


\bibliographystyle{aa} 
\bibliography{mybib}

\label{lastpage}

\end{document}